\documentstyle[12pt,psfig]{article}
\newcommand{\lb}{\bibitem}
\begin{document}
\renewcommand{\thefootnote}{\fnsymbol{footnote}}
\begin{center}
{\LARGE {\bf Spectral Representation at Finite Temperature}}\vspace{.5in}\\
{\large Sourav Sarkar$^a$, B. K. Patra$^a$, V. J. Menon$^b$, and S. Mallik$^c$}
\vspace{.3in}\\
{\it $^a$ Variable Energy Cyclotron Centre, 1/AF Bidhan Nagar, Calcutta 700 064,
India\\
$^b$ Department of Physics, Banaras Hindu University, Varanasi 221 005, India\\
$^c$ Saha Institute of Nuclear Physics, 1/AF Bidhan Nagar, Calcutta 700 064, 
India\\}
\vspace{.4in}

\underline{\large Abstract} 
\end{center}
\addtolength{\baselineskip}{0.5\baselineskip}
This is a short review on the thermal, spectral representation in the real-time 
version of the finite temperature quantum field theory. After presenting
a clear derivation of the spectral representation, we discuss
the properties of its spectral function. Two applications of this 
representation are then considered. One is the solution of the Dyson equation 
for the thermal propagator. The other is the formulation of the QCD sum rules 
at finite temperature.
\vspace{.2in}

\pagestyle{myheadings}
\markright{}
\section{Introduction}
\setcounter{equation}{0}
\renewcommand{\theequation}{2.\arabic{equation}}

The real- and the imaginary- time versions of the quantum field theory
in a medium (i.e., at finite temperature and/or density) have somewhat
complementary virtues~\cite{landsman}. The real-time version is closer to the 
conventional (vacuum) field theory, involving no sum over frequencies and so 
requiring no analytic continuation to real energies. However, the price to 
pay for this closeness is the $2 \times 2$ matrix structure of all two-point 
correlation functions. Generally speaking, static thermodynamic quantities 
of a system are usually calculated in the imaginary-time version, while the 
real-time version appears convenient for calculating more detailed, 
particularly time dependent, quantities.

Here we are concerned with the real-time spectral representation for
the thermal, two-point correlation functions of local operators. (If this
operator is chosen to be just a field operator, the representation gives
the corresponding thermal propagator.) Although 
the real-time version is of rather recent development, such thermal spectral 
representations were obtained by Landau~\cite{landau} as early as 1958, after 
K\"{a}llen and Lehmann~\cite{kallen} derived them for the propagators in 
the vacuum. Earlier Low~\cite{low} had derived representations 
for non-vacuum matrix elements in the time component $q_0$ at fixed space 
component $\vec{q}$ of the $4$-momentum variable $q_\mu$, conjugate to the 
coordinate difference $x^\mu$ of the two operators in the matrix element. The
Landau representation is of this variety, extending such matrix elements
to their ensemble average.

Of course, Landau's derivation of the spectral representation in real-time
was technically incomplete, as he did not take into account its $2 \times 2$
matrix structure. The complete representation was written by Semenoff
and Umezawa~\cite{semenoff} in 1983, after Umezawa and his 
collaborators~\cite{umezawa} had established the real-time version.

In this work we derive in detail the spectral representation for the 
two-point correlation functions of local operators, obtaining the symmetry 
relations satisfied by the spectral function. To keep the kinematics simple
we consider Lorentz scalar operators. As an example we calculate the spectral
function to the leading order in scalar field theory. We 
go on to consider two applications of this spectral representation. One is to
review the well-known reduction of the self-energy matrix to essentially
a single function by using its factorized structure~\cite{kobes}. The other is to
formulate the QCD sum rules in the real time, finite temperature field
theory~\cite{bochkarev}.

The derivation of the spectral representation is given in sec.2, giving an 
example of calculation of the spectral function in sec.3. In sec.4 we 
consider two applications, one to Dyson equation and the other to QCD sum 
rules. Finally in sec.5 we present a summary of the results derived
in the paper.

\section{Spectral Representation}
\setcounter{equation}{0}
\renewcommand{\theequation}{2.\arabic{equation}}

We choose the contour in the complex time plane as originally proposed 
by Umezawa~\cite{umezawa}, to get a symmetric $2 \times 2$ matrix for the 
free thermal propagator. It consists effectively of two segments, one running 
along the real axis in the positive direction and the other parallel to it, but shifted by 
$-i\beta/2$, in the reverse direction, where $\beta$ is the inverse 
temperature $T$. 

Consider the contour-ordered, two-point correlation function of a 
local operator. This operator may, in general, be  a composite one,
built out of fundamental fields, which may be scalar, spinor or
vector fields. Thus in QCD the composite operator may be any one of the
conserved currents, namely the vector current, $\bar q (x) \gamma_\mu
\frac{\tau_i}{2} q (x)$ or the axial vector current,
 $\bar q (x) \gamma_\mu \gamma_5 \frac{\tau_i}{2} q (x)$, where $q(x)$
is the doublet field of $u$ and $d$ quarks and $\tau_i$'s are the
Pauli matrices. Besides such bosonic operators, one may also
have fermionic operators like the so-called baryon currents. For the proton 
this operator is $\epsilon^{abc} [ {u^a}^T (x) C \gamma_\mu u^b (x)] 
{[ \gamma^\mu \gamma^5 d^c (x) ]}_D$, where  $a$, $b$, $c$ are the colour 
indices, $C$ is the charge conjugation matrix and $D$ a Dirac 
index~\cite{ioffe}. If the operator is not a composite one, but just one of 
the fundamental fields themselves, the correlation function becomes the 
thermal propagator for that field.

To avoid kinematic complications, we choose in the following
a Lorentz scalar operator, which may be composite or fundamental and
denote it by $O(x)$. Further we assume it to be bosonic. Then
the contour-ordered two-point function may be put in the
form of a $2 \times 2$ matrix,
\begin{equation}
{\cal T} (x-y) = i\left( \begin{array}{cc} \langle T O(x) O(y) \rangle & 
\langle O(y-i\beta/2) O(x) \rangle \\
\langle O(x-i\beta/2) O(y) \rangle & \langle \overline{T} O(x-
i\beta/2) O(y-i\beta/2) \rangle \end{array} \right)~,
\label{full matrix}
\end{equation} 
where $T$ and $\overline{T}$ denote the usual time and anti-time ordering 
and $\langle \cdot \cdot \cdot \rangle$ denotes the ensemble average; thus 
for the operator $O$,
\begin{eqnarray}
\langle O \rangle = Tr~O e^{-\beta H}/Z,~~~~Z=Tr~e^{-\beta H}~,
\end{eqnarray}
$H$ being the Hamiltonian of the system and $Tr$ denoting trace over any 
complete set of states. In momentum space, the Fourier transform is denoted 
by the same symbol,
\begin{equation}
{\cal T}_{ab}(q) = \int d^4z e^{iqz}~{\cal T}_{ab}(z),~~~~a,b=1,2~.
\label{fourier}
\end{equation}

Let us first consider the $11$-component of the matrix. We evaluate the 
trace over a complete set of states $|m \rangle $, $m=1,2,..$, which are 
eigenstates of the $4$-momentum operator $P_\mu$ with eigenvalues ${(p_m)}_\mu$.
Denoting ${(p_m)}_0$ by $E_m$, it becomes
\begin{equation}
\langle T O(x) O(y) \rangle = Z^{-1}\sum_m e^{-\beta E_m} \langle m|T O(x) 
O(y)| m \rangle ~,
\label{11xy}
\end{equation}
which is a sum over forward amplitudes weighted by the corresponding
Boltzmann factors. Again inserting the same set of states to extract the
co-ordinate dependence of the field operators, we get
\begin{eqnarray}
\langle T O(z) O(0) \rangle &=& Z^{-1} \sum_{m,n} e^{-\beta E_m} 
\left( 
\theta (z^0) e^{i(p_m-p_n)\cdot z} + \theta (-z^0) e^{-i(p_m-p_n)\cdot z} 
\right) \nonumber\\
&&\times {|\langle m| O(0)|n \rangle |}^2 ~.
\label{11z}
\end{eqnarray}
It is now simple to work out the Fourier transform (\ref{fourier}). The
integration over space gives rise to $\delta$-functions in $3$-momentum, while 
that over the time variable produces the energy denominators. Inserting a 
$\delta$-function in the energy variables, we may put it in the form
\begin{equation}
{\cal T}_{11}(q) = -\frac{1}{2\pi} \int_{-\infty}^{\infty} d q_0^\prime
\left( \frac{M^+(q_0^\prime, \vec{q})}{q_0-q_0^\prime +i \epsilon} -
 \frac{M^-(q_0^\prime, \vec{q})}{q_0-q_0^\prime -i \epsilon} 
\right)~,
\label{D11q}
\end{equation}
where
\begin{equation}
M^\pm(q_0,\vec{q})= Z^{-1}\sum_{m,n}e^{-\beta E_m} {(2\pi)}^4 \delta
(q\pm p_m\mp p_n){| \langle m| O(0)|n\rangle|}^2
\label{Mpmdoub}
\end{equation}
The double sum over states may be converted back to the product of field 
operators giving
\begin{equation}
M^+(q)= \int d^4z e^{iqz} \langle O(x) O(0) \rangle~,
\label{M+q}
\end{equation}
and $M^-(q)$ having an identical expression with $O (x)$ and $O (0)$
interchanged.

In the case of the vacuum expectation value, the two spectral functions, 
which are functions of $q^2$ only, can be shown to be equal by the use of the 
causality requirement. However, in the present case, where they are functions 
of $q_0$ and $|\vec{q}|$ (or $q^2$ and $u \cdot q$ in a Lorentz covariant 
framework with $u^\mu$ being the four-velocity of the medium), a
similar argument does not go through. But we still have two relations
connecting $M^\pm$, namely the Kubo-Martin-Schwinger relation in 
momentum space,
\begin{equation}
M^+(q_\mu) = e^{\beta q_0} M^-(q_\mu)~,
\label{kms}
\end{equation}
and the symmetry relation,
\begin{equation}
M^+(-q_\mu) = M^-(q_\mu)~.
\label{msymm}
\end{equation}
These relations are usually obtained from the operator representation
(\ref{M+q}). They may also be obtained from the double sum 
representation (\ref{Mpmdoub}) : the relation (\ref{msymm}) is evident,
while to get the relation (\ref{kms}) we interchange the dummy
indices $m,n$ in any one of $M^\pm (q)$ and use the $\delta$-function to 
express $E_m-E_n$ by $q_0$. 

Let us now define the spectral function $\rho$ as
\begin{equation}
\rho(q_0,\vec{q}) \equiv M^+(q_0,\vec{q}) - M^-(q_0,\vec{q}) \nonumber\\
= \int d^4 z e^{iqz} \langle [O (x), O(0) ] \rangle~,
\label{rhoq}
\end{equation}
which, on noting (\ref{msymm}), is antisymmetric under $q_\mu \rightarrow - 
q_\mu$,
\begin{equation}
\rho (-q_\mu) = - \rho(q_\mu)~.
\label{rhoanti}
\end{equation}
Also using (\ref{kms}), we can express both $M^\pm$ in terms of $\rho$
\begin{eqnarray}
M^+(q_\mu) = \frac{e^{\beta q_0}}{e^{\beta q_0} -1} \rho(q_\mu)~, \nonumber\\
M^-(q_\mu) = \frac{1}{e^{\beta q_0} -1} \rho(q_\mu)~. 
\label{M+M-}
\end{eqnarray}
\par We now wish to redefine the energy denominators in (\ref{D11q}) with the
Feynman $i\epsilon$ prescription. For this purpose we write it as,
\begin{eqnarray}
{\cal T}_{11}(q) &=&i\int_{-\infty}^{\infty} \frac{dq_0^\prime}{2\pi}
\left[
\frac{i\rho(q_0^\prime,\vec{q})}{q_0-q_0^\prime
+iq_0^\prime \epsilon}+ \pi \delta(q_0-q_0^\prime)\right. \nonumber\\ 
&& \times\left.\{ M^+(q_0^\prime) +
M^-(q_0^\prime)- {\mathrm {sgn}} (q_0^\prime) (M^+(q_0^\prime) - 
M^-(q_0^\prime)) \} \frac{}{}
\right]
\end{eqnarray}
where the $\vec{q}$ dependence of $M^\pm$ is suppressed. Here we have written
the integrals in (\ref{D11q}) first as their principal values and then with the
indicated $i\epsilon$ prescription, the terms with $\delta$-functions serving
to compensate these changes. Folding the range of integration on to $(0,\infty)$ and using the 
relations (\ref{rhoanti}) and (\ref{M+M-}), we get the desired result \footnote
{The change of sign of $\vec{q}$ while using (2.12) is of no consequence here,
since it occurs either as $\vec{q}\cdot \vec{q}$ or $\vec{q} \cdot \vec{p}$, 
where $\vec{p}$ is a $3$-vector to be integrated out over its entire range.},
\begin{equation}
{\cal T}_{11}(q) =i\int_{0}^{\infty} \frac{d{q_0^\prime}^2}{2\pi}
\rho(q_0^\prime,\vec{q}) \left[ \frac{i}{{q_0}^2-{q_0^\prime}^2
+i\epsilon} + 2 \pi \delta({q_0^\prime}^2 - {q_0}^2) \frac{1}{
e^{\beta |q_0|}-1} \right]
\label{finalD11q}
\end{equation}
\par The $22$-element of the matrix ${\cal T}$ may be simplified by
invoking the translational invariance, 
\begin{eqnarray}
{\cal T}_{22}(x-y) =  \theta(x^0-y^0) \langle O(y) O(x)\rangle +
\theta(y^0-x^0) \langle O(x) O(y) \rangle
\end{eqnarray}
Repeating the steps similar to above, we get for ${\cal T}_{22}(q)$ an 
expression identical to the one for ${\cal T}_{11}(q)$, except for 
complex conjugation of its first term in square bracket,
\begin{equation}
{\cal T}_{22}(q) =i\int_{0}^{\infty} \frac{d{q_0^\prime}^2}{2\pi}
\rho(q_0^\prime,\vec{q}) \left[ \frac{-i}{{q_0}^2-{q_0^\prime}^2
-i\epsilon} + 2 \pi \delta({q_0^\prime}^2 - {q_0}^2) \frac{1}{
e^{\beta |q_0|}-1} \right]
\end{equation}
The $12$- and $21$- elements turn out to be identical,
\begin{eqnarray}
{\cal T}_{12}(q) = {\cal T}_{21}(q)&=& i\frac{e^{\beta q_0/2}}
{e^{\beta q_0}-1} \rho(q_0,\vec{q})\nonumber\\
&=& i \int d {q_0^\prime}^2 \rho(q_0^\prime,\vec{q}) \delta ({q_0^\prime}^2 -
q_0^2) \frac{e^{\beta |q_0|/2}}{e^{\beta |q_0|}-1}
\end{eqnarray}
Recognizing the density distribution function and the
free propagator in vacuum with Feynman boundary condition,
\[
n= \frac{1}{e^{\beta |q_0|}-1}~,~~~~ {\Delta} =\frac{i}{q_0^2-{q_0^\prime}^2
+i \epsilon}
\]
in the above expressions, we collect the results for the components as
\begin{equation}
{\cal T}_{ab}(q_0,\vec{q}) =\int_{0}^{\infty} \frac{d{q_0^\prime}^2}{2\pi}
\rho(q_0^\prime,\vec{q}) {\cal D}^0_{ab}(q_0,q_0^\prime)~,
\label{Dabq}
\end{equation}
where ${\cal D}^0_{ab}$ is the free thermal propagator,
\begin{eqnarray}
{\cal D}^0_{ab} & =&i\left( \begin{array}{cc} (1+n) {\Delta} +n
{\Delta}^\ast & 
\sqrt {n(1+n)} ({\Delta}+ {\Delta}^\ast) \\
\sqrt {n(1+n)}({\Delta} + {\Delta}^\ast) & n{\Delta} +(1+n)
{\Delta}^\ast
\end{array} \right) 
\label{D0ab}\\
&=& U(|q_0|)~i\left( \begin{array}{cc} {\Delta} & 0 \\
0 & {\Delta}^\ast 
\end{array} \right) U(|q_0|)~,
\label{d0diag}
\end{eqnarray}
with
\begin{eqnarray}
U  = \left( \begin{array}{cc} \sqrt{1+n} & \sqrt{n} \\
\sqrt{n} & \sqrt{1+n} 
\end{array} \right)~. 
\end{eqnarray}
The matrix $U$ does not depend on the integration variable in (\ref{Dabq}) 
and like the free propagator, the correlation function also factorizes,
\begin{eqnarray}
{\cal T}_{ab}(q_0,\vec{q})=U(|q_0|) \left( \begin{array}{cc} T & 0 \\
0 & -T^\ast \end{array} \right) U(|q_0|)
\label{Dab}
\end{eqnarray}
where 
\begin{eqnarray}
T(q_0,\vec{q})  = \int_{0}^{\infty} \frac{d{q_0^\prime}^2}{2\pi}
\frac{\rho(q_0^\prime,\vec{q})}{q_0^{\prime 2} - q_0^2 -i \epsilon} 
\end{eqnarray}
Thus the matrix function is given essentially by a single analytic function 
$T(q_0, \vec{q})$. This integral representation, as such, may not converge, 
when it would need subtractions. However, this point does not concern us in the
following.

As a simple example, let us calculate $T(q_0, \vec{q})$ for the
case where the operator $O(x)$ is a scalar field operator $\phi(x)$,
representing particles of mass $m$. Then from Eq.(2.11) we get 
immediately the spectral function as

\[
\rho(q_0,\vec{q}) = 2 \pi~{\mathrm {sgn}} (q_0) \delta(q_0^2 -{\vec{q}}~^2
-m^2)
\]
giving
\begin{eqnarray}
T(q) =\frac{-1}{q^2-m^2+i\epsilon}
\label{rho0}
\end{eqnarray}
which is the free propagator function for the scalar field with
Feynman boundary conditions.\footnote{The occurrence of $-1$ instead
of $i$ in the numerator of Eq. (2.25) is due to the presence of $i$
in the definition of (2.1) of the correlation function.}

The important point to notice here is that there is only one spectral
function $\rho$ giving all the four components of the correlation
function and it may be obtained by calculating any one of the
components, say ${\cal T}_{11}$. But although $\rho \equiv (M^+-M^-)$
is (twice) the imaginary part of $T$, it is not so for ${\cal T}_{11}$ : As 
seen from (\ref{D11q}), (twice) the imaginary part of ${\cal T}_{11}$ is 
$(M^+ + M^-)$. The two, however, are related through eq.(\ref{M+M-}),
\begin{eqnarray}
\rho(q_0,\vec{q}) =2  \tanh(\beta q_0/2)~Im {\cal T}_{11} (q_0,\vec{q}) 
\label{tanh}
\end{eqnarray}
Their real parts are, however, equal,
\begin{eqnarray}
Re T(q_0,\vec{q}) = Re {\cal T}_{11} (q_0,\vec{q})~.
\end{eqnarray}

Eqs.(2.23 - 24) constitute the finite temperature extension of the
K\"{a}llan-Lehmann representation in vacuum. The breakdown of Lorentz
invariance is explicit as it treats $\vec{q}$ as fixed. (As already mentioned,
it can be formally
restored, but only by introducing the four-velocity of the heat bath.) Note
the relations (2.26-27) between (the real and imaginary parts of) $T$
and the ${\cal{T}}_{11}$- component. Such relations also exist among the components of
the self-energy matrix, to be discussed below. 

Before leaving this section, we wish to point out that an equation
like (2.7) giving the spectral function as a double sum over
states is not very convenient for its evaluation. For, the $\delta$-function
in it seems to indicate that even initial and final states with
a large number $m$ and $n$ of particles would give a contribution
in the low energy region provided their momentum difference
$(p_m-p_n)$ is small. Actually most of these particles are from
the heat bath and do not participate in the interaction with
the operators in the two point function. Thus it is much simpler
to calculate the spectral function from the relevant Feynman
diagrams, where the thermal propagator properly takes into
account the particles of the heat bath in the distribution
function present in it. Accordingly, we turn to the Feynman diagrams
in the next section to calculate the spectral function.

\section{An Example }
\setcounter{equation}{0}
\renewcommand{\theequation}{3.\arabic{equation}}

Let us evaluate the correlation function perturbatively for the case where
\begin{eqnarray}
O(x) = \phi_1(x) \phi_2(x) \quad ,
\end{eqnarray}
the fields $\phi_1(x)$ and  $\phi_2(x)$ being two different scalar
fields representing particles of masses $m_1$ and $m_2$. Then the
perturbative evaluation for the $11$-component follows from
\begin{eqnarray}
T_{11}(q) = i \int d^4 x~e^{iqx}~\langle \frac{}{} T e^{i \int 
{\cal{L}}_{int} (y)
d^4 y} \phi_1 (x) \phi_2 (x) \phi_1 (0) \phi_2 (0)  \frac{}{}\rangle ~,
\end{eqnarray}
where ${\cal{L}}_{int} (\phi_1, \phi_2)$ is the interaction part of the
Lagrangian involving $\phi_1(x)$ and $\phi_2(x)$ (Fig.1). Usual
Feynman rules apply with the vacuum propagators being replaced by the
thermal ones,
\begin{eqnarray}
\int d^4 x e^{iqx} \langle \frac{}{} T \phi_i (x) \phi_i(0) 
\frac{}{} \rangle =
\frac{i}{k^2 -m_i^2} + 2 \pi \delta(k^2 -m_i^2) n_i (k), \quad i=1,2
\end{eqnarray}
where $n_i(k)$ is the distribution function of particle type $i$. To
leading order the $11$- component is given by
\begin{eqnarray}
{\cal D}_{11}(q) = -i \int \frac{d^4k}{{(2\pi)}^4} \{ (1+n_1)\Delta_1
+n_1\Delta_1^\ast \} \{ (1+n_2)\Delta_2 + n_2 \Delta_2^\ast \}
\label{D11qex}
\end{eqnarray}
where
\begin{eqnarray}
&& n_1 = \frac{1}{e^{\beta w_1}-1},~~~~ w_1 = \sqrt{\vec{k}^2+m_1^2} \nonumber\\
&& n_2 = \frac{1}{e^{\beta w_2}-1},~~~~ w_2 = \sqrt{{(\vec{k}-\vec{q})}^2+m_2^2} \nonumber\\
&& \Delta_1 = \frac{i}{k^2-m_1^2+i\epsilon},~~~~\Delta_2 = \frac{i}{{(k-q)}^2-m_2^2 + i\epsilon}
\end{eqnarray}

%%%%%%%%%%%%%%%%%%%%%%%%%%%%%%%%%%%%%%%%%%%%%%%%%%%%%%%%%%%%%^M
\begin{figure}
\psfig{file=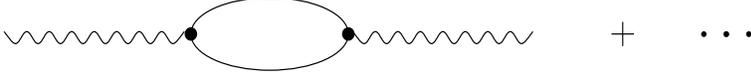,height=1cm,width=10cm}
\caption{Two point function for $O (x) = \phi_1(x) \phi_2(x)$. The
dots indicate higher order diagrams, whose nature and 
contribution depend on ${\cal{L}}_{int} (x)$
}
\end{figure}

%%%%%%%%%%%%%%%%%%%%%%%%%%%%%%%%%%%%%%%%%%%%%%%%%%%%%%%%%%%%%%%%%%%

Each term in the the integrand of (\ref{D11qex}) has a product of two propagators. Their
singularities in $k_0$ are only due to the poles in these propagators. So
the integration over $k_0$ is performed by closing the contour in 
the $k_o-$plane \footnote{By power counting the integral in (3.2) is divergent.
But the divergence resides only in the real part.}, after which the imaginary 
part may easily be read off,
\begin{eqnarray}
Im {\cal D}_{11} &=& \pi  \int \frac{d^3 \vec{k}}{{(2\pi)}^3}\,\, 
\frac{1}{4 w_1 w_2}\nonumber\\
&&\left[\frac{}{} \{ (1+n_1)(1+n_2) +n_1n_2 \} \{\delta(q_0-w_1-w_2)+ 
\delta(q_0+w_1+w_2) 
\} \right. \nonumber\\ 
&& + \left. \{(1+n_1)n_2+(1+n_2)n_1 \} \{ \delta(q_0-w_1+w_2)+
\delta(q_0+w_1-w_2) 
\}\frac{}{} \right]\nonumber\\
\label{imd11}
\end{eqnarray}

It will be noted that the factors involving the density distribution
functions are not appropriate for the interpretation in terms of emission
and absorption probabilities of the particles~\cite{weldon}. The
desired interpretation follows, if we use
the energy conserving $\delta$-functions
to rewrite it as
\begin{eqnarray}
Im {\cal D}_{11} = \coth(\beta q_0/2)~I~, 
\label{coth}
\end{eqnarray}
where
\begin{eqnarray}
I(q_0,\vec{q}) &=& \pi  \int \frac{d^3 \vec{k}}{{(2\pi)}^3 4 w_1 w_2}
\left[\frac{}{} (1+n_1+n_2) \{\delta(q_0-w_1-w_2)-\delta(q_0+w_1+w_2) \}
\right. \nonumber\\
&&+ \left. (n_2-n_1) \{ \delta(q_0-w_1+w_2)-\delta(q_0+w_1-w_2) 
\}\frac{}{} \right]~.
\end{eqnarray}
Thus from (2.26) and (3.5) we see that the $\tanh$- and 
$\coth$- factors cancel out in the spectral function giving
\begin{eqnarray}
\rho(q_0,\vec{q}~) =2 I (q_0,\vec{q}~)~,
\end{eqnarray}
which agrees with the one obtained from the imaginary-time 
formulation~\cite{weldon}. Also notice that $I(q_0,\vec{q}~)$ is 
antisymmetric under $q_0 \rightarrow -q_0$ in agreement with Eq.(2.12).

Although we are discussing in this work only the bosonic propagator
and bosonic intermediate states in its spectral representation, 
the cancellation of trigonometric functions is quite general.
As long as the
propagator is bosonic, there arises a $\coth$- factor in the spectral function,
even if the intermediate state is a fermion-antifermion system. But if we 
consider a fermionic propagator, the $\tanh$- and $\coth$- factors interchange
in the expressions analogous to (\ref{tanh}) and (\ref{coth}). Thus the 
trigonometric factors cancel in the spectral representation
in all cases. Also we note that if there is a chemical
potential $\mu$ in the Boltzmann factor, the argument, $\beta q_0/2$ of the 
trigonometric factors are replaced by $\beta (q_0-\mu)/2$, besides
other changes.\vspace{.05in}\\ 

\section{Applications}
\setcounter{equation}{0}
\renewcommand{\theequation}{4.\arabic{equation}}
\vskip 0.1in
\noindent {\bf A. Dyson Equation}
\vskip 0.1in

The spectral representation is very general in that it is valid for all
$q^2$. If, however, we are interested only in the immediate neighbourhood
of the pole in the propagator, the Dyson equation is the appropriate
tool to work with (Fig.2). But even here the factorizibility of the full
propagator, as established by the spectral representation, may be
used to show that the self-energy matrix appearing in the Dyson equation
has actually a simple structure, reducing essentially to a single function.

At finite temperature the Dyson equation for the propagator matrix in 
momentum space is

%%%%%%%%%%%%%%%%%%%%%%%%%%%%%%%%%%%%%%%%%%%%%%%%%%%%%%%%%%%%%
\begin{figure}
\psfig{file=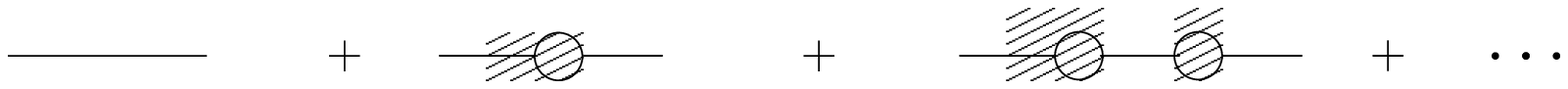,height=0.61cm,width=14cm}
\caption{The series of diagrams summed up in the Dyson equation for the 
propagator}
\end{figure}

%%%%%%%%%%%%%%%%%%%%%%%%%%%%%%%%%%%%%%%%%%%%%%%%%%%%%%%%%%%%%%%%%%%

\begin{eqnarray}
{\cal D} = {\cal D}_0 - {\cal D}_0 \Sigma {\cal D}~,
\label{dyson}
\end{eqnarray}
where $\Sigma$ is the self-energy matrix. It has the
solution
\begin{eqnarray}
{\cal D}^{-1} = {\cal D}_0^{-1} + \Sigma 
\label{caldin}
\end{eqnarray}
Using the factorized forms for the propagators, it gives
\begin{eqnarray}
\left( \begin{array}{cc} 1/D & 0 \\ 0 & -1/D^\ast
\end{array} \right) = \left( \begin{array}{cc} -(q^2-m^2) & 0\\ 0 & q^2-m^2 
\end{array} \right) + U \Sigma U
\label{factorization}
\end{eqnarray}
It follows that $\Sigma$ must be of the form
\begin{eqnarray}
\Sigma(q)= U^{-1}(q_0) \left( \begin{array}{cc} \tilde{\Sigma} & 0 \\ 0 & 
-\tilde{\Sigma}^\ast \end{array} \right) U^{-1}(q_0) 
\end{eqnarray}
where $\tilde{\Sigma}(q)$ is the self-energy function. Then we have the
solution 
\begin{eqnarray}
D =- \frac{1}{q^2-m^2-\tilde{\Sigma}+i\epsilon}
\end{eqnarray}
All the components of the self-energy matrix can now be expressed through the
function $\tilde{\Sigma}$. From (4.4) we get~\cite{kobes} 
\begin{eqnarray}
&&\Sigma_{11}=-\Sigma_{22}^\ast=Re \tilde{\Sigma} + i (1+2n) Im \tilde{\Sigma} 
\nonumber\\
&&\Sigma_{12}=\Sigma_{21} = -2i \sqrt {n(1+n)} Im \tilde{\Sigma}
\end{eqnarray}

These relations should not be interpreted as constraints imposed on the
elements of the $\Sigma$-matrix by the solution of the Dyson equation. They 
are, in fact, automatically satisfied by the expressions obtained from
perturbation theory. The above method of  solution is just a convenient way
to establish them.

Let us note here that the factorization (\ref{factorization}) of the 
{\em full} matrix
propagator is really not necessary to solve the Dyson
equation. Indeed, writing $\Sigma=U^{-1}\Sigma^\prime 
U^{-1}$ and using the factorization of the {\em free} propagator alone, eqn.(\ref{caldin})
gives
\begin{eqnarray}
{\cal D} = U {\left( \begin{array}{cc} - (q^2-m^2) + {\Sigma^\prime}_{11} &
{\Sigma^\prime}_{12} \\ {\Sigma^\prime}_{21} & (q^2-m^2)+{\Sigma^\prime}_{22}
\end{array} \right)}^{-1} U
\end{eqnarray}
On obtaining the inverse, one finds that each of the components of the 
propagator matrix is a linear combination of the pole terms, ${(q^2-m^2-\Sigma_+)}^{-1}$ and 
${(q^2-m^2-\Sigma_-)}^{-1}$, where
\begin{eqnarray}
\Sigma_\pm = \frac{1}{2} (\Sigma^\prime_{11} - \Sigma^\prime_{22})
\pm \frac{1}{2} { \left( {(\Sigma^\prime_{11} + \Sigma^\prime_{22})}^2
-4 \Sigma^\prime_{12} \Sigma^\prime_{21} \right) }^{1/2} ~.
\end{eqnarray}
Thus although we can solve Dyson's equation, we cannot relate the components 
of $\Sigma$ to a single function
$\tilde{\Sigma}$, without using factorizibility of the propagator.
But on using these relations, we see
that $\Sigma_\pm$ indeed reduces respectively to
$\tilde{\Sigma}$ and ${\tilde{\Sigma}}^\ast$.

The two-particle contribution to the spectral function considered
in sec.3 may also be taken as a contribution to the self-energy. Thus
the calculation of ${\cal{D}}_{11}$ there also serves as an example of
$\Sigma_{11}$.

\vskip 0.3in
\noindent {\bf B. Thermal QCD Sum Rules}
\vskip 0.1in

A vacuum QCD sum rule may be derived for a two-point correlation function, 
$ \langle 0 | T {\cal{O}}_1(x) {\cal{O}}_2 (0) | 0 \rangle$
of any two operators ${\cal{O}}_1(x)$ and ${\cal{O}}_2(x)$, which can
be built out of quark and gluon fields of the QCD theory. As already
mentioned, these operators can be mesonic, like the conserved currents of 
the flavour $SU(2)$ group, or can be fermionic, like the baryon currents. 
The QCD sum rules~\cite{svz} are obtained by equating the Fourier transforms 
of the spectral representation to that of the Operator Product Expansion 
(OPE) of the two-point function at a convenient space-like momentum.

These vacuum sum rules have already been extended to the medium at finite 
temperature (with or without chemical potential)~\cite{bochkarev,hatsuda}. 
It is not our purpose here to review this topic in detail. Instead, we merely 
show how one deals with the $2\times 2$ matrix structure of the two-point
function in the formulation of these sum rules. 

To bring out the essential features of these thermal QCD sum rules
without getting involved in kinematic complications, we 
again consider the two-point correlation function (2.1) of a bosonic
Lorentz scalar operator $O(x)$, 
satisfying the spectral representation (2.23). Further, if we take 
$q^2$ space-like, $q^2 = q_0^2 - {\vec{q}}^{~2} = -Q^2$, $Q^2 > 0$, 
where the function $T(Q^2, {\vec{q}}^{~2})$ is real, it reduces to
\begin{eqnarray}
{\cal T}_{ab}(Q^2, {\vec{q}}^{~2}) = T (Q^2,{\vec{q}}^{~2}) \left( 
\begin{array}{cc} 1 & 0 \\
0 & -1 \end{array} \right) 
\label{Dabsumrule}
\end{eqnarray}
where
\begin{eqnarray}
T(Q^2,{\vec{q}}^{~2}) = \int_{-{\vec{q}}^{~2}}^{\infty} \frac{dq^2}{2\pi}
\frac{\rho(q^2,{\vec{q}}^{~2})}{q^2 + Q^2}
\label{specsumrule}
\end{eqnarray}
with $\rho$ given by (2.26). But as we already pointed out at the
end of sec.2, it is more useful to calculate it perturbatively
within the context of an appropriate effective field theory.

It is now clear that the $11$- (or $22$-) component of the matrix
(4.9) contains all the information in the space-like region. We may
then obtain the thermal spectral sum rules by considering this component
alone. 

The other ingredient needed to write the sum rules is the OPE for the 
$11$-component. It is in deriving this expansion that the actual quark
and/or gluon structure of the operator $O(x)$ must be
spelt out. What the OPE does is to expand the two-point
function in terms of local operators having finite
matrix elements, its singular behaviour at short distance
being segregated in $c$ number (Wilson) coefficient functions.\footnote{
The off-diagonal elements are actually regular as the two
arguments of the operator ${\cal{O}}$ cannot tend to each other
for finite $\beta$.} At sufficiently high $Q^2$, its Fourier transform
may be written as, 
\begin{eqnarray}
i \int d^4 x e^{i qx} \langle T {{O}}(x) {{O}}(0) \rangle_{11} = \sum  c_i 
\frac{ \langle {{O}}_i \rangle}{{(Q^2)}^m}~,
\label{ope}
\end{eqnarray}
where $c_i$ are numbers. The local operators ${{O}}_i$ are labelled by the
index $i$ denoting their dimensions. The inverse powers of $Q^2$
are determined by dimensional counting, $m=i/2 + 2 -p$,
where $p$ is the dimension of the operator ${{O}}(x)$.

The sum rule follows from equating the spectral representation
(4.10) to the OPE (4.11) in a convenient region of $Q^2$. After Borel
improvement~\cite{svz}, it finally becomes
\begin{eqnarray}
\frac{1}{2\pi M^2} \int_{-{|\vec{q}|}^2}^\infty d s~e^{-s/M^2} 
\rho(s, {\vec{q}}^2) = \sum  c_i \frac{ \langle {{O}}_i \rangle}{(m-1)!{(M^2)}^m}
~,
\end{eqnarray}
where $M$ is the Borel mass replacing $Q$.

In writing sum rules of practical interest, we have, of course, to deal with 
more complicated kinematics  depending on the nature of the
operators ${{O}}(x)$ in the two-point function. But it is clear that the
method of derivation and the structure of these sum rules will be the same
as in the above illustrative example. The special features of these
sum rules over the corresponding vacuum ones and their evaluations are
described in the literature~\cite{hatsuda,mallik1,mallik2}.

\section{Conclusion}

We present a detailed derivation of the spectral representation and
discuss properties of the spectral function. We calculate 
the spectral function, as a proto-type example, in a scalar field
theory and note its agreement with the expression
obtained from the imaginary time version of the finite temperature field 
theory. Relationships similar to those among the components of the
self-energies are shown to exist also among the components of the two-point
correlation matrix.

We then review two important applications of this representation. One is 
the solution of the Dyson equation for the thermal
propagator. The
factorizability of the full propagator (as well as of the free one) leads
to a simple solution, relating at the same time all the four components of the
self-energy matrix to a single function. 
We also point out that the Dyson equation could be solved even without
this input of factorizability, but then the simple form of the self-energy
matrix could not be exposed. The latter could, of course,
be inferred from its perturbative calculation in each order.

The other application relates to the formulation
of the QCD sum rules in the real time thermal field
theory. We show how the matrix structure of the two-point function
simplifies at space-like momenta, where it is needed for writing the
sum rules. We can then 
work with its $11$- component, obtaining the familiar
sum rules written in the literature.

%\newpage

\end{document}